\documentclass[aps,prd,a4paper,nofootinbib,
preprintnumbers,superscriptaddress,showpacs,showkeys,twocolumn]{revtex4-2}

\usepackage{amsmath}
\usepackage{amssymb}
\usepackage{bm}
\usepackage{graphicx}
\usepackage{xcolor}
\usepackage{slashed}
\usepackage{comment}
\usepackage[normalem]{ulem}

\usepackage[sort,compress]{cleveref}

\crefname{section}{Sec.\!}{Secs.\!}
\crefname{equation}{Eq.\!}{Eqs.\!}
\crefname{figure}{Fig.\!}{Figs.\!}
\crefname{table}{Tab.\!}{Tabs.\!}
\crefname{appendix}{App.\!}{Apps.\!}
\crefname{chapter}{Chapter}{Chapters}

\crefname{section}{Sec.\!}{Secs.\!}
\crefname{equation}{Eq.\!}{Eqs.\!}
\crefname{figure}{Fig.\!}{Figs.\!}
\crefname{table}{Tab.\!}{Tabs.\!}
\crefname{appendix}{App.\!}{Apps.\!}

\newcommand{\be}{\begin{equation}}
\newcommand{\ee}{\end{equation}}
\newcommand{\ba}{\begin{eqnarray}}
\newcommand{\ea}{\end{eqnarray}}

\begin{document}

\title{Topological Susceptibility in the Superconductive Phases of Quantum Chromodynamics: a Dyson-Schwinger Perspective}

\author{Fabrizio Murgana}\email{fabrizio.murgana@dfa.unict.it}
\affiliation{Department of Physics and Astronomy "Ettore Majorana", University of Catania, Via Santa Sofia 64, I-95123 Catania, Italy}\affiliation{INFN-Sezione di Catania, Via Santa Sofia 64, I-95123 Catania, Italy}

\author{Giorgio Comitini}\email{giorgio.comitini@dfa.unict.it}
\affiliation{Department of Physics and Astronomy "Ettore Majorana", University of Catania, Via Santa Sofia 64, I-95123 Catania, Italy}\affiliation{INFN-Sezione di Catania, Via Santa Sofia 64, I-95123 Catania, Italy}

\author{Marco Ruggieri}\email{marco.ruggieri@dfa.unict.it}
\affiliation{Department of Physics and Astronomy "Ettore Majorana", University of Catania, Via Santa Sofia 64, I-95123 Catania, Italy}\affiliation{INFN-Sezione di Catania, Via Santa Sofia 64, I-95123 Catania, Italy}


\begin{abstract}
We 
test non-perturbative gluon propagators 
recently studied in the literature,
by computing
the topological susceptibility, $\chi$, of 
the 
superconductive phases of 
Quantum Chromodynamics at high density. We formulate the problem within the High-Density Effective Theory,
and use the 2-particle irreducible formalism
to compute the effective potential of the
dense phases. 
We focus on superconductive phases
with
two and three massless
flavors.
Within this formalism,
we write a Dyson-Schwinger
equation in the rainbow approximation
for the anomalous part of the 
quark propagator in the
superconductive phases,  
in which the non-perturbative
gluon propagator plays its role.
We complete the model by adding  
a $U(1)_A$-breaking term whose coupling
is fixed perturbatively at large quark
chemical potential.
We then  use the effective potential to compute
$\chi$ in the superconductive phases.
We finally discuss implications 
of the results for the axion mass in 
superdense phases of Quantum Chromodynamics.
\end{abstract}



\maketitle

\section{Introduction}

The theory of strong interactions, 
Quantum Chromodynamics (QCD), 
can be augmented  by a term that does not break
gauge invariance, but which is odd under parity
as well as under the combined operations of parity and charge
conjugation.
This term is topological in nature and can be written as
\begin{equation}
S_\theta = \theta \int d^4x \, q(x),
\label{eq:Stheta}
\end{equation}
where
$\theta$ is a real number sometimes called the $\theta$-angle
of QCD,
$q(x)$ is the topological charge density, 
\begin{equation}
q(x) = \frac{g^2}{64\pi^2} \epsilon^{\mu\nu\rho\sigma} F^a_{\mu\nu} F^a_{\rho\sigma},
\label{eq:topological_charge_density}
\end{equation}
and $F^a_{\mu\nu}$ denotes the field strength tensor
of the gluon field.
Through a Fujikawa transformation, the $\theta$-term can be reabsorbed into the quark sector, introducing an anomalous phase in the quark mass term which ultimately leads to the explicit breaking of the $ U(1)_A $ symmetry. 
In the framework of the instanton liquid model, this breaking is associated with an effective multi-fermion interaction among quarks due to instanton exchange. 

In the context of QCD at finite $\theta$, it is useful to
introduce
the  topological susceptibility, $\chi$, 
defined as the second derivative of the effective
potential of QCD with respect to $\theta$ at 
$\theta=0$. $\chi$ quantifies 
the fluctuations of the
topological charge in QCD. This quantity has been extensively studied in the vacuum and at finite temperature, both within effective models \cite{Zhang:2023lij,Lu:2018ukl,Gatto:2011wc,Ruggieri:2020qtq,Aoki:2017imx}, as well as with $\chi$PT \cite{Mao:2009sy,GrillidiCortona:2015jxo,Landini:2019eck,Bottaro:2020dqh} and in Lattice QCD simulations \cite{Borsanyi:2016ksw, Bonati:2015vqz,Bernard:2012fw}. However, studies of $ \chi $ at finite chemical potential remain scarce, mainly due to the sign problem in Lattice QCD. In particular, very little is known about the behavior of topological fluctuations in the color-superconducting phase of QCD, where the formation of diquark condensates and the modification of the axial anomaly could significantly impact the topological sector of the theory.

In this work,
we compute $\chi$ 
in superdense phases of QCD, in which the system could be
in a color-superconductive phase \cite{Alford:1998mk,Rajagopal:2000wf, Nardulli:2001iv,Nardulli:2002ma,Schafer:1999jg, Buballa:2003qv, Blaschke:2005uj,Alford:2007xm, Shovkovy:2004me,Shovkovy:1999mr,Gatto:2007ja,Son:1998uk,Abuki:2001be}. 
We make use of  a renormalized version
of QCD around the Fermi surface of quarks, known 
as High Density Effective Theory (HDET),
see~\cite{Nardulli:2002ma} for a review.
In our  work, we combine HDET 
with a non-local 
Dyson-Schwinger equation
(DSE) for the quark propagator in the rainbow approximation
(bare vertices)
supplemented with
improved gluon propagators~\cite{Comitini:2024xjh}.
We derive the relevant DSE
within the 2-particle-irreducible (2PI)
formalism of Cornwall-Jackiw-Toumbulis (CJT)~\cite{Cornwall:1974vz}.
This requires the formulation of dense QCD at finite $\theta$,
which has not been studied within the aforementioned
approach yet.
The propagators of~\cite{Comitini:2024xjh} 
have been used so far only to get a semi-quantitative 
picture of the QCD phase diagram, but have not been used
in 
calculations of QCD properties 
like condensates, susceptibilities and so on. 
Therefore, with this work
we offer a first test of their impact on the QCD condensates
and on the QCD thermodynamic potential.

Several studies have been carried out in the DSE framework for the color superconductive phases, see for example \cite{Hong:1999fh,Muller:2013pya,Muller:2016fdr,Malekzadeh:2006ud,Hou:2004bn,Schafer:1999jg}. Compared to these studies, in this work we 
write a DSE for the quark propagator within the HDET formalism,
and use an improved gluon propagator which was recently obtained in~\cite{Comitini:2024xjh},
that entails non-perturbative screening masses.

In order to compute the topological susceptibility, we include an explicit $U_A(1)$ breaking term~\cite{T'Hooft:76, T'Hooft:86} which effectively describes the one-instanton exchange (OIE) among quarks, as well as their coupling to the
$\theta-$angle when a standard Fujikawa rotation is performed
to remove the parity-odd term~\eqref{eq:Stheta} 
from the QCD Lagrangian.
This term has already been  used in effective models
with a contact interaction~\cite{Zhang:2023lij,Ruggieri:2020qtq, Gatto:2011wc,Murgana:2024djt,Lu:2018ukl,Zhang:2025lan,Balkin:2020dsr, Zhang:2025lan}, 
in combination with an effective coupling describing
the one-gluon exchange (OGE). 
In this work, rather than embracing the
approach of local effective models,
we start with the full QCD lagrangian, 
renormalize it around the Fermi surface isolating only
the relevant terms at large $\mu$, then
we add the OIE at finite $\theta$. 
The dependence of 
the coupling in the OIE channel, $\zeta$,
on
the quark chemical potential, $\mu$, and the QCD coupling,
$g$, is fixed by its perturbative expression at large $\mu$,
both in the two-flavor and in the three-flavor models we consider.

One of the potential applications of this work is the
study of the QCD-axion 
\cite{Peccei:1977hh, Peccei:1977ur, Wilczek:1977pj, Balkin:2020dsr} in dense QCD phases.
In fact, the axion potential is directly accessible
from the QCD potential at finite $\theta$
by the formal replacement $\theta\rightarrow a/f_a$ where
$a$ denotes the axion field and 
$f_a$ is the axion decay constant. Hence,
the calculation of the effective QCD potential at finite
$\theta$ allows us to access the low-energy properties
of the axion, like the mass, the self-coupling and so on.
We will comment briefly on the impact of our results
on the axion mass in Section VII, leaving
a more complete study of this interesting problem to
future works.

The plan of the article is as follows. In Section II
we summarize the HDET and the 2PI formalisms adapted to the
CFL phase, as well as the gluon propagator we use in the
DSE. In Section III we derive the HDET gap equation
within the HDET. In Section IV we present the gap equation
for the CFL phase, and set the basic relations that help to
write the topological susceptibility in a compact form.
In Section V we discuss our calculation of the topological
susceptibility of the CFL phase. In Section VI we briefly
discuss the topological suceptibility of the 2SC phase.
In Section VII we discuss semi-analytical relations
between $\chi$ and the couplings of the model in 
two limiting cases.
In Section VIII we comment on the relevance of our results
for the computation of the QCD axion mass in 
color-superconductive phases.
Finally, in Section IX we present our conclusions and an
outlook. We use $\hbar=c=1$ throughout the article.

\section{Model and Calculation Setup}

\subsection{HDET  for color-superconductive phases}

The High-Density Effective Theory (HDET) is a very well established framework for the study of strongly interacting matter under conditions of extremely high density \cite{Nardulli:2002ma,Casalbuoni:2001de,Hong:1998tn,Hong:1999ru}. It provides a systematic method to explore these regimes by focusing on the degrees of freedom most relevant at high baryon density, significantly simplifying the calculations but still effectively capturing the high-density behavior of the theory. 
The HDET has been used to study the properties of color-superconducting phases, allowing for semi-analytical calculations of the gap parameters and critical temperatures for various pairing patterns~\cite{Murgana:2024djt, Casalbuoni:2001ha,Casalbuoni:2001de, Beane:2000ms,Anglani:2011cw,Casalbuoni:2002pa, Nardulli:2001iv, Casalbuoni:2001gt,Gatto:2007ja,Anglani:2013gfu,Anglani:2006br,Casalbuoni:2004wm,Casalbuoni:2003sa,Casalbuoni:2002my,Casalbuoni:2001gt,Casalbuoni:2002pa}. Recently, it has also been used to construct an effective model to describe the quark-condensation pattern on the
 Fermi surface \cite{Jeong:2024rst}. By focusing on the Fermi surface dynamics, HDET can also provide  insight into the equation of state (EoS) of dense matter, a critical ingredient for modeling neutron stars and other compact astrophysical objects \cite{Nardulli:2002ma,Casalbuoni:2001de}.
Furthermore, HDET has been used to elucidate the properties of collective excitations, such as Nambu-Goldstone bosons arising from symmetry breaking in dense QCD phases \cite{Gatto:2007ja}.

The HDET of QCD is well known, therefore we limit ourselves
to summarize the main ideas of the formulation;
for more details, we refer to~\cite{Nardulli:2002ma}.
The key idea behind HDET is that at $T = 0$, the vacuum is characterized by fermions filling all available low-energy states up to the Fermi energy. Due to the Pauli exclusion principle, interactions involving low-energy quarks necessarily involve the exchange of high-momentum particles. This is suppressed in QCD because of asymptotic freedom. Consequently, 
the leading contribution to the thermodynamic properties
of dense QCD comes from
the quarks near the Fermi surface. Thus, the effective Lagrangian in HDET is constructed by expanding around the Fermi surface and including only terms that are relevant close to the Fermi surface. The resulting theory describes quasi-particles and their interactions, thus providing a framework to analyze the pairing mechanisms and gap structures.

In the HDET, the quark momenta are decomposed as
\begin{equation}
    p^\mu = \mu v^\mu + \ell^\mu,
\end{equation}
where $\mu$ denotes the quark chemical potential, and 
$v^\mu = (0, \bm{v})$ with $\bm{v}$ is the Fermi velocity versor, $|\bm{v}| = 1$. $\ell^\mu$ is called the residual
momentum, whose spatial part is
\begin{equation}
\bm{\ell} = \bm{v} \ell_\parallel + \bm{\ell}_\perp, 
\end{equation}
Here, 
$\bm\ell_\parallel$ and $\bm\ell_\perp$ correspond to the
parallel and perperndicular components of the residual
momentum.
However, one can always choose the velocity parallel to $\bm{p}$, so that $\bm{\ell}_\perp = 0$.
The $4$-integral measure of the HDET reads
\begin{equation}\label{eq:ufficio_1421}
    \int \frac{d^4p}{(2\pi)^4}  = 
    \frac{4\pi\mu^2}{(2\pi)^4} \sum_{\bm{v}} \int_{-\infty}^{+\infty}d\ell_\parallel \int_{-\infty}^{+\infty} d\ell_0. 
\end{equation}
with
\begin{equation}
    \sum_{\bm{v}} \equiv \int \frac{d\bm{v}}{8\pi} =
   \frac{1}{2}\int\frac{d\phi_v d\theta_v}{4\pi},
   \label{eq:misura_angolare_yyy}
\end{equation}
where 
$(\phi_v,\theta_v)$ correspond to the azimuthal and 
polar angles of the Fermi velocity, and 
an additional factor $1/2$ is included in order to avoid double counting.

The leading-order HDET lagrangian density is
\begin{equation}
\mathcal{L}_{QCD}^{HDET}=\mathcal{L}_g+\mathcal{L}_D,
\label{eq:sempreliasinistracontheoeleao}
\end{equation}
where
\begin{equation}
\mathcal{L}_g=  -\frac{1}{4}F^{a}_{\mu\nu}F^{a\, \mu\nu}
\end{equation}
is the standard gluon term, and 
\begin{equation}
\mathcal{L}_D=\mathcal{L}_0+\mathcal{L}_1
\label{eq:dir}
\end{equation}
is the quark Lagrangian. 
In  \cref{eq:dir}, $\mathcal{L}_0$ represents the free Dirac Lagrangian, which in our framework can be written as
\begin{equation}
\mathcal{L}_0=  \sum_{\vec{v}} \left( \psi_+^{\dagger} iV \cdot \partial \psi_+ + \psi_-^{\dagger} i\tilde{V} \cdot \partial \psi_- \right)
\end{equation}
while $\mathcal{L}_1$ represents the interaction term with the gluon field $A$,
\begin{equation}
\mathcal{L}_1 = i g \sum_{\bm{v}} \left( \psi_+^{\dagger} iV \cdot A \psi_+ + \psi_-^{\dagger} i\tilde{V} \cdot A \psi_- \right)\label{eq:inzaghinonlemandaAdire}
\end{equation}
Here $V=(1, \bm{v})$, $\tilde{V}=(1, -\bm{v})$,  with $\bm{v}$ being the aforementioned quark Fermi velocity, and $\psi_\pm$ indicate the positive-energy velocity-dependent fields with opposite velocity
\begin{equation}
    \psi_\pm\equiv\psi_{+,\,\pm\bm{v}}.
\end{equation}

We include an additional term, $\mathcal{L}_4$, in the Lagrangian density~\eqref{eq:sempreliasinistracontheoeleao}, that explicitly breaks $U(1)_A$ symmetry and describes the interaction among quarks due to the
one-instanton exchange.
Following~\cite{Murgana:2024djt}, we take 
\begin{eqnarray}
\mathcal{L}_4 &=& -\zeta \left[
(\psi_L^T i C \psi_L)
(\psi^\dagger_R i C \psi^*_R)e^{i\theta}\right.
\nonumber\\
~~~&+&\left.
(\psi_R^T i C \psi_R)
(\psi^\dagger_L i C \psi^*_L)e^{-i\theta}
\right],\label{eq:tandemst2}
\end{eqnarray}
where $\theta$ corresponds to the $\theta$-angle of QCD,
$L$ and $R$ denote left- and right-handed quark fields.
The lagrangian~\eqref{eq:tandemst2}
is obtained from the one-instanton-exchange
effective
interaction at $\theta=0$~\cite{T'Hooft:76, T'Hooft:86,Schafer:1996wv} via
a global $U(1)_A$ rotation that removes 
the gluonic $\theta$-term~\eqref{eq:Stheta}
and encodes all the information 
about $\theta$
into the quark sector. 
The interaction~\eqref{eq:tandemst2}
has already been used in order to study the coupling 
of the QCD-axion to quarks in a color-superconductive phase \cite{Murgana:2024djt, Zhang:2025lan,Hell:2009by} and, with suitable modifications, also in the framework of chiral symmetry breaking \cite{Zhang:23, Lopes:2022efy,Bandyopadhyay:2019pml,Lu:2018ukl, Mohapatra:2022wvj,Das:2020pjg,Gong:2024cwc,Kumar:2024abb,Bersini:2025yvt}, including applications to the study of domain walls \cite{Workman:22, Gabadadze:2000vw,Davidson:1984ik, Takahashi:2018tdu,Davidson:1983tp}. 
For the three-flavor model that we mostly consider in 
our study, the interaction~\eqref{eq:tandemst2}
has to be understood as an effective interaction term
in the limit of small quark masses: in this limit,
the original six-fermion term becomes effectively a
four-quark interaction~\cite{Schafer:2002ty}.
We make this explicit in the definition of the 
coupling $\zeta$, see below.

The coupling in the interaction~\eqref{eq:tandemst2}
is unknown, unless one considers
very large values of $\mu$ where perturbative calculations
are meaningful.
In this asymptotic limit, 
assuming a diagonal quark mass matrix
$\mathcal{M}=\mathrm{diag}(M_u,M_d,M_s)$
and putting $M_u = M_d = M_s \equiv M$,
for the three-flavor model we would have~\cite{Kruglov1990, Schrempp:2005vc,Shuryak:2021fsu, Shifman:1979uw,Schafer:1996wv,Son:2000fh,Schafer:2002ty}
\begin{equation}
\zeta= M \int d\rho \, n_0(\rho)  \frac{ 2 N_c-1 }{4 \left(N_c^2-1\right) (2 N_c)} 2 (2 \pi  \rho )^4 \rho ^3.
\label{eq:zeta_coupl_jjj}
\end{equation}
Here $n_0$ denotes the instanton density, 
\begin{equation}
n_0(\rho) = C_N \left(\frac{8\pi^2}{g^2}\right)^ {\left( 2N_c \right) }
\rho^{-5} \exp\left(-\frac{8\pi^2}{g^2}\right) 
e^{-N_f \mu^2 \rho^2}, 
\end{equation}
where 
$\rho$ is the instanton size. Moreover,
$N_f$ is the number of quark flavors, $g$ the QCD coupling (that we assume as independent of the instanton size), $N_c$ is the number of colors and $C_{N_c}$ is a coefficient given by 
 \begin{equation}
 C_{N_c} = 0.466 e^{-1.679 N_c} \frac{1.34^{N_f}}{(N_c - 1)!(N_c - 2)!}\,.   
\end{equation}
Using a constant $g$ and $N_f=N_c=3$ we get
\begin{equation}
    \zeta=6.106\times 10^9\times
    \left(\frac{1}{g}\right)^{10}
     e^{-\frac{8 \pi ^2}{g^2}}
    \frac{1}{g^{2} \mu ^2}\frac{M}{\mu}. 
    \label{eq:theaboveequation}
\end{equation}
Given the uncertainties on the values of $g$ and $M$
at finite $\mu$, we treat $\zeta$ as a parameter in our
model, and for the sake of simplicity we assume that
its dependence on  $\mu$,  $M$ and $g$ is the one 
given by Eq.~\eqref{eq:theaboveequation}.
Hence, 
for a given value of $\mu$
we let $g$ and $M$  vary, and fix $\zeta$ 
via Eq.~\eqref{eq:theaboveequation}.
If mass degeneracy is broken then instead of $M$
one should use $\mathrm{Tr}\mathcal{M}/3$
in Eq.~\eqref{eq:zeta_coupl_jjj}.

In the CFL phase, the $SU(3)_c$ color symmetry is completely broken, and so is the chiral $SU(3)_A\otimes SU(3)_V$ flavor symmetry with a left $SU(3)_{c+V}$ that combines color and flavor. The ansatz for the quark-quark condensate
in the phase is \cite{Alford:1998mk}
\begin{equation}
\langle\psi^{LT}_{\alpha i}C\psi^{L}_{\beta j}\rangle=-\langle\psi^{RT}_{\alpha i}C\psi^{R}_{\beta j}\rangle
\propto
\frac{\Delta}{2}\epsilon_{\alpha\beta I}\epsilon_{i j I}.
\label{eq:calciodirigorenondato}
\end{equation}
Here, $\alpha,\beta$ are color indices,
$i,j$ are flavor indices, $C=i\sigma_2$, and 
a sum over $I=1,2,3$ is understood. 
$\Delta$ is the color-superconductive gap parameter, 
that sets the scale
for the gap in the spectrum of the quark quasi-particles.
It is useful to introduce the CFL color-flavor basis
\begin{equation}
       \psi_{+,\alpha, i}=\sum_{A=1}^9\frac{\lambda_{A_\alpha,i}}{\sqrt{2}}\psi_{+,A},\label{eq:moltariserva_1101}
   \end{equation}
where $\lambda_A$ with $A=1,\dots,8$ denote the set of 
Gell-Mann matrices, normalized as $\mathrm{Tr}(\lambda_A\lambda_B) = 2\delta_{AB}$, and
$\lambda_9=\lambda_0=\sqrt{\tfrac{2}{3}}\mathbb{I}$. 
In this basis, the gap matrix reads~\cite{Nardulli:2002ma}
\begin{equation}
\Delta_{AB} = \Delta \mathrm{Tr}(\varepsilon_I T_A^T \varepsilon_I T_B)=\Delta_A\delta_{AB},\label{eq:gapmatrix1}
\end{equation}
with
\begin{equation}
\Delta_A =\left\{\begin{array}{cl}
\Delta,  &A=1\cdots 8,\\
-2\Delta,  &A=9,
\end{array}\right.\label{eq:gapmatrix2}
\end{equation}
where the first 8 modes are degenerate in an octet and the ninth corresponds to a singlet. 
This split of the quark spectrum in the CFL phase
remains also in case an additional color-sextet
contribution is included, see~\cite{Shovkovy:1999mr,Rajagopal:2000wf, Alford:1998mk,Schafer:2000tw, Buballa:2003qv}; 
for the sake of simplicity,
we neglect the 
condensation in the 
color-sextet channel since its contribution is generally much
smaller than that of the anti-triplet 
interaction~\cite{Casalbuoni:2003cs}.

We introduce the Nambu-Gorkov spinor basis as 
\begin{equation}
    \chi_A=\left(\begin{array}{c}
         \psi_{+,A} \\
          C\psi^*_{-,A}
    \end{array}\right),
\end{equation}
with $C=i\sigma_2$.
In this basis, the quark-gluon vertex 
in Eq.~\eqref{eq:inzaghinonlemandaAdire} 
can be expressed as: 
\begin{equation}
\Gamma_\mu^{aAB}=ig\left(\begin{array}{cc}
         V_\mu h_{AaB}&0  \\
        0 & -\tilde{V}h^*_{AaB}
    \end{array}\right),\label{eq:gitadifamiglia}
\end{equation}
with 
\begin{equation}
    h_{AaB}=\mbox{Tr }\left[T_AT_aT_B\right], 
\end{equation}
$A,B=1,\cdots9$, $a=1,\cdots 8$ and
$T_A= \lambda_A/\sqrt{2}$. 
The quark-quark interaction mediated by the exchange of 
one gluon, with vertices~\eqref{eq:gitadifamiglia},
is dubbed the one-gluon-exchange (OGE).

In the CFL basis,  $\mathcal{L}_4$  in Eq.~\eqref{eq:tandemst2} reads
\begin{eqnarray}
\mathcal{L}_4 &=& 
\zeta\Gamma_{ABCD}
\left[
(\psi_{LA}^T C \psi_{LB})
(\psi^\dagger_{RC}  C \psi^*_{RD})e^{i\theta}
\right.\nonumber\\
&&+\left.
(\psi_{RA}^T  C \psi_{RB})
(\psi^\dagger_{LC}  C \psi^*_{LD})e^{-i\theta}
\right],
\label{eq:gamma_ABCD_eq}
\end{eqnarray}
with
\begin{equation}
\Gamma_{ABCD} = \Gamma_{AB}\Gamma_{CD},~~~\Gamma_{AB}=
\mathrm{Tr}(\varepsilon_I T_A^T \varepsilon_I T_B).
\label{eq:gamma_def_ABCD}
\end{equation}
$\Gamma_{AB}=\Gamma_A\delta_{AB}$ 
has the same color-flavor structure of the
gap matrix, see Eqs.~\eqref{eq:gapmatrix1} and~\eqref{eq:gapmatrix2}, that is
\begin{equation}
\Gamma_A =\left\{\begin{array}{cl}
1, &A=1\cdots 8,\\
-2, &A=9.
\end{array}\right.\label{eq:gapmatrix3}
\end{equation}

\begin{widetext}

\subsection{Ansatz on the full quark propagator}

In order to define an HDET-2PI effective potential for the
CFL phase, we need an ansatz on the full quark propagator,
$S$; this is a matrix in color-flavor and Nambu-Gorkov spaces.
Similarly to previous works, we adopt the following form
for $S^{-1}$ for left- and right-handed fields:
\begin{equation}\label{eq:prop}
 S_{AB}^{-1}(\ell)=\left( \begin{array}{cc}
      V\cdot \ell  & \Delta_A(\ell) \\
       \Delta_A(\ell) & \tilde{V}\cdot\ell
   \end{array} \right) \delta_{AB}.
\end{equation}
Analogously, the inverse free propagator is
\begin{equation}\label{eq:propAAA}
  S_{0, AB}^{-1}(\ell)=\left( \begin{array}{cc}
      V\cdot \ell  & 0 \\
       0 & \tilde{V}\cdot\ell
   \end{array} \right) \delta_{AB}.
\end{equation}
The inversion of \cref{eq:prop} is trivial and leads to 
\begin{equation}
S_{AB} = S_A\delta_{AB},\label{eq:aldottorato}
\end{equation}
with
\begin{equation}\label{eq:s}
    S_{A}(\ell)=\frac{1}{V\cdot\ell\,\tilde{V}\cdot\ell-\Delta_A^2(\ell)+i\varepsilon}\left(\begin{array}{cc}
       \tilde{V}\cdot\ell & -\Delta_A(\ell)\\
       -\Delta_A(\ell)& V\cdot\ell
    \end{array}\right).
\end{equation}
We will use the ansatz~\eqref{eq:s} for both left- and
right-handed fields in this work.

The off-diagonal component of the inverse quark propagator,
$\Delta_A(\ell)$, will be computed by solving the self-consistent
gap equation in the 2PI formalism, see Sections~\ref{sec:muoverlabene} and~\ref{sec:ilmilandipioli}.
The interactions that contribute to $\Delta$ in our work
are the OGE and the OIE interactions.
As a final comment, we remark that in this work
we consider the static limit,
and consider homogeneous and isotropic condensates only,
therefore
the quark self-energy depends only on $\ell_\parallel$.

\subsection{2PI Effective Potential}

The Cornwall-Jackiw-Tomboulis (CJT) formalism \cite{Cornwall:1974vz} is a widely used tool in quantum field theory, particularly in contexts where non-perturbative effects and self-consistent treatment of quantum fluctuations play a critical role, such as dynamical symmetry breaking \cite{Pilaftsis:2015bbs,Aarts:2002dj,Tsutsui:2017uzd,Wetterich:2002ky}, out-of-equilibrium dynamics \cite{Calzetta:2008iqa,Berges:2015kfa,Aarts:2002dj}, critical and strongly correlated systems  \cite{Blaizot:2021ikl,Berges:2015kfa,Dupuis:2005ij} and stochastic processes \cite{Bode_2022}.
Within this formalism,  the concept of the  1PI-effective action \cite{Goldstone:1962es, Jona-Lasinio:1964zvf}, which depends on the expectation value of the quantum fields, is improved considering a 2PI-effective action, depending both on the 1-point functions (the field expectation values) and the 2-point functions, the propagators.


Hence,
the CJT-effective potential 
that we use
can be expressed as 
\begin{equation}
    V(S)=-i\int\frac{d^4 p}{(2\pi)^4}\mbox{ Tr}\left[\ln{S_0^{-1}(p)S}(p)-S_0^{-1}(p)S(p)+I\right]+ \frac{i}{2}\int\frac{d^4 p}{(2\pi)^4}\mbox{ Tr}\left[\ln{D_0^{-1}(p)D(p)}-D_0^{-1}(p)D(p)+I\right]+V_2(S),
    \label{eq:VS_1_lla}
\end{equation}
where $S$ and $D$ corresponds to the full quark and gluon propagators, respectively, while $S_0$ and $D_0$ correspond to the bare ones.
$V_2(S)$ corresponds to the sum of all the 2PI vacuum diagrams which are produced by the interaction vertices of the theory, and include
the coupling of the quarks to the gluons
via the one-gluon-exchange and the 
one-instanton-exchange. 
In this work, gluons are dynamical particles, since they have a propagator; however,   we do not compute the gluon propagator self-consistently via a DSE coupled to the one for the quark propagator: instead, we use the results obtained non-perturbatively from a different approach \cite{Comitini:2024xjh} that we describe in more detail in the next section. Therefore, the second addendum in the r.h.s. of \cref{eq:VS_1_lla} is merely a constant and we neglect it in the calculation of the   effective potential, and does not contribute to the DSE of the quark propagator. Hence, the CJT potential we consider is  
\begin{equation}
    V(S)=-i\int\frac{d^4 p}{(2\pi)^4}\mbox{ Tr}\left[\ln{S_0^{-1}(p)S}(p)-S_0^{-1}(p)S(p)+I\right]+ V_2(S).
    \label{eq:VS_1_lla2}
\end{equation}

\begin{figure}[t!]
    \centering
    \includegraphics[width=0.5\linewidth]{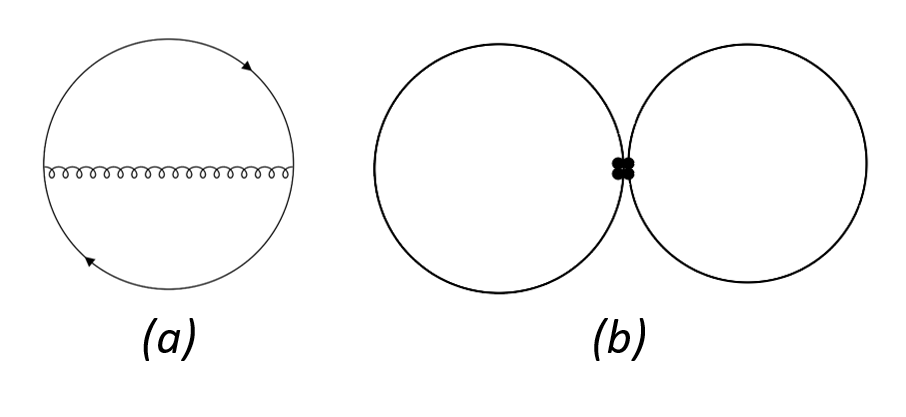}
    \caption{Two-loop diagrams contributing to $V_2$ in the
    CJT potential in Eq.~\eqref{eq:ufficio_1056}.
    Diagrams (a) and  (b) correspond to the OGE and the OIE contact term respectively.}
    \label{Fig:V2}
\end{figure}

Taking into account OGE and OIE
terms, we can write $V_2$ as
\begin{equation}
V_2 = V_2^{\mathrm{OGE}} + V_2^{\mathrm{OIE}},
\label{eq:ufficio_1056}
\end{equation}
where
\begin{equation}
V_2^{\mathrm{OGE}}= \mathrm{Tr}[\Gamma S\Gamma D S],
\label{eq:V2_one_g}
\end{equation}
and
\begin{equation}
V_2^{\mathrm{OIE}}=\mathrm{Tr} [S \Gamma^t S].
\label{eq:V2_t}
\end{equation}
Here, 
$\Gamma$ and $\Gamma^{t}$ denote the bare 
vertices of the OGE and OIE interactions
respectively.
Using the CFL basis~\eqref{eq:moltariserva_1101}
we can write the contributions to $V_2$ as 
\begin{equation}
V_2^{\mathrm{OGE}}=\frac{1}{2}\int \frac{d^4 p}{(2\pi)^4} \frac{d^4k}{(2\pi)^4}\, \mathrm{Tr}\left[\Gamma_\mu^{aAC}S_{CD}(p)\Gamma_\nu^{bDB}D_{\mu\nu}^{ab}(p-k) S_{AB}(k) \right]+ (L\rightarrow R),
\label{eq:geometriaelementare_1103}
\end{equation}
and 
\begin{equation}
   V_2^\mathrm{OIE}= \frac{1}{2}\zeta\int \frac{d^4 p}{(2\pi)^4} \frac{d^4k}{(2\pi)^4}\, \mathrm{Tr}\left[S_{AB}(\ell)\Gamma_{ABCD}S_{CD}(k)e^{i\theta} \right]\,+h.c.\label{eq:ipotenusaecateti}
\end{equation}
A diagrammatic representation of~\eqref{eq:ufficio_1056}
is shown in Fig.~\ref{Fig:V2}, where
(a) and (b) respectively denote the contributions
from the OGE and the OIE.

\end{widetext}

\subsection{Improved gluon Propagator}

We now 
describe the gluon propagator we use in our calculations.
At $\mu\neq0$, as a consequence of the breaking of Lorentz symmetry, the gluon propagator cannot be expressed in terms of a single scalar function. Instead, it is determined by two functions $D^{T}(|{\bm p}|)$ and $D^{L}(|{\bm p}|)$ which enter the gluon propagator $D_{\mu\nu}^{ab}({\bm p})$ evaluated at $p^{0}=0$ as
\begin{equation}\label{tensprop}
    D_{\mu\nu}^{ab}({\bm p})=-\delta^{ab}\left[D^{T}(|{\bm p}|)\,P_{\mu\nu}^{T}(p)+D^{L}(|{\bm p}|)\,P_{\mu\nu}^{L}(p)\right]\ .
\end{equation}
In the Landau gauge, assuming that the medium has four-velocity $u^{\mu}=\delta^{\mu}_{0}$, the projectors $P^{T}(p)$ and $P^{L}(p)$ take the form
\begin{eqnarray}
P_{\mu\nu}^{T}(p) &=& (1-\delta_{\mu0})(1-\delta_{\nu0})\left(g_{\mu\nu}+\frac{p_{\mu}p_{\nu}}{|{\bm p}|^{2}}\right),\\
P_{\mu\nu}^{L}(p) &=& t_{\mu\nu}(p)-P_{\mu\nu}^{T}(p),
\end{eqnarray}
where $t_{\mu\nu}(p)=g_{\mu\nu}-p_{\mu}p_{\nu}/p^{2}$ is the usual four-dimensionally transverse projector. They satisfy
the chain of equalities
\begin{eqnarray}
&&P^{T/L}(p)\cdot P^{T/L}(p)=P^{T/L}(p),\\
&&P^{T}(p)\cdot P^{L}(p)=0,\\
&&P^{T/L}(p)\cdot t(p)=t(p)\cdot P^{T/L}(p)=P^{T/L}(p),
\end{eqnarray}
are both orthogonal to the four-vector $p$, $P^{T/L}(p)\cdot p = 0$, and 
in the static limit $p^{0}=0$ they further simplify as
\begin{eqnarray}
&&P_{\mu0}^{T}(p)=P_{0\mu}^{T}(p)=0,~~~P_{ij}^{T}(p)=-\delta_{ij}+\frac{p_{i}p_{j}}{|{\bf p}|^{2}}, \\
&&P_{00}^{L}(p)=1,~~~P_{\mu i}^{L}(p)=P_{i\mu}^{L}(p)=0.
\end{eqnarray}
Since $P^{T}(p)$ is orthogonal to the spatial projection $\overline{p}=(0, {\bm p})$ of $p$, $P^{T}(p)\cdot \overline{p} = 0$, we will refer to the former as the (three-dimensionally) transverse projector, and to $P^{L}(p)$ as the (three-dimensionally) longitudinal one; accordingly, we will call $D^{T}(|{\bm p}|)$ and $D^{L}(|{\bm p}|)$ the transverse and the longitudinal component of the gluon propagator.

The evaluation of $D^{T}(|{\bm p}|)$ and $D^{L}(|{\bm p}|)$ is essentially a non-perturbative task. Since for $p^{0}=0$ one has $p^{2}=-|{\bm p}|^{2}<0$, the components of the gluon propagator could in principle be extracted from the lattice data in Euclidean space. However, at non-vanishing chemical potentials, the lattice is plagued by the infamous sign problem, which limits the applicability of unquenched Monte Carlo methods to domains where the quark determinant is positive, such as QCD at finite isospin density or 2-color QCD at finite baryonic density and an even number of quark flavors. Recent calculations carried out in these domains \cite{BBNR20,BR21,BNRT21} suggest that the transverse component of the gluon propagator is only weakly dependent on chemical potential\footnote{At least for not too large chemical potentials $\mu\lesssim800$~MeV, corresponding to $a\mu\ll1$ (with $a$ the lattice spacing), where the lattice simulations are more reliable.}, whereas the longitudinal one gets strongly suppressed as chemical potential increases, a behavior which is consistent with a Debye mass proportional to chemical potential being generated by the medium in the longitudinal sector. As $\mu\to 0$, SO(4) rotational invariance -- the equivalent of Lorentz invariance in Euclidean space -- is restored, and the two components collapse to a unique function which remains finite in the deep infrared, signaling that a dynamical mass is generated for the gluons already in the vacuum by a mechanism which is independent of the medium. This phenomenon has been the subject of numerous studies over the last decades \cite{LSWP98a,AN04,AP06,BIMS09,TW10,PTW14,DOS16,RSTW17,NAHP21}, is hypothesized to be caused by gluonic condensates \cite{DGSVV08,DOV10,CDDS24} and is nowadays considered an established feature of the gluon propagator.

\begin{figure}[t!]
    \centering
    \includegraphics[width=\linewidth]{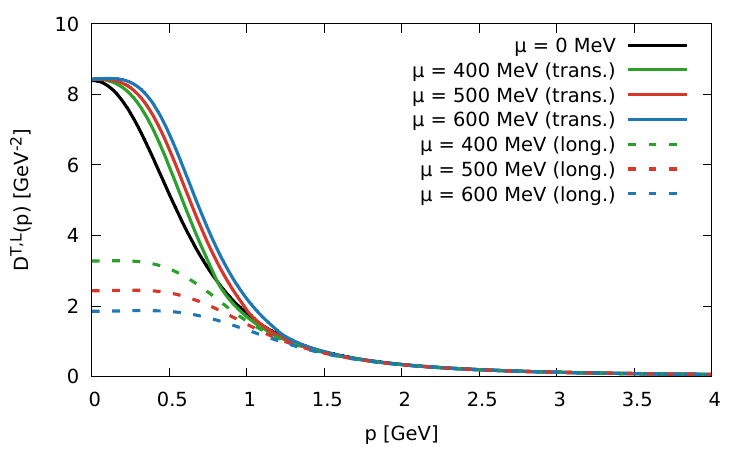}
    \caption{Transverse and longitudinal components of the gluon propagator $D^{T/L}$ as a function of 
    the $3-$momentum $p=|\bm p|$, for different values of the chemical potential $\mu$.} 
    \label{fig:gluprops}
\end{figure}

Results that display the same behavior as the lattice were recently obtained \cite{Comitini:2024xjh} for the full QCD gluon propagator at finite baryonic chemical potential using a perturbative method known as the screened massive expansion. The latter \cite{SIR16a,SIR16b,SC18,CS20} consists in a redefinition of the QCD perturbative series carried out in such a way that the (four-dimensionally) transverse gluons propagate as massive already at tree level, while leaving the action of QCD unchanged; the non-perturbative content of the expansion is condensed in a gluon mass parameter $m$ which must be provided as an external input, e.g. by setting the energy units of the propagator itself. In Fig.~\ref{fig:gluprops} we show the propagator computed within the screened massive expansion  of full QCD at $\mu=400,500,600$~MeV for $N_{f}=2+1$ quarks with masses equal to the up/down and strange current masses. The free parameters of the expansion were fixed by extracting the mass scale of the optimized propagator at $\mu=0$ from pure Yang-Mills lattice data like in \cite{SC18,Comitini:2024xjh}, which yields a gluon mass parameter $m=656$~MeV. While the screened massive expansion is able to provide expressions for the propagator which, at finite chemical potential, are analytic up to a one-dimensional integral, in what follows it will be useful to parametrize the components of the propagator using the Gribov-Stingl form
\begin{equation}\label{gribov-stingl}
    D^{T/L}(x)=\frac{Z^{T/L}(x^{2}+(m_{1}^{T/L})^{2})}{x^{4}+2(m_{2}^{T/L})^{2}x^{2}+(m_{3}^{T/L})^{4}}\ ,
\end{equation}
where $Z^{T/L}$, $m_{1,2,3}^{T/L}$ are free parameters, different for each component and dependent on chemical potential. Their values, as obtained by fitting the propagators in Fig.~\ref{fig:gluprops}, are reported in Tab.~\ref{tab:params}; the resulting fits are essentially indistinguishable from the propagators obtained by a numerical integration.
In the DSE we will make use
of the
static gluon propagator~\eqref{gribov-stingl} with
$x=|\bm \ell - \bm p|$, namely
\begin{equation}
\label{propagatore:comitini}
     D^{T/L}(\bm{\ell}-\bm{p})=\frac{Z^{T/L}(|\bm{\ell}-\bm{p}|^{2}+(m_{1}^{T/L})^{2})}{|\bm{\ell}-\bm{p}|^{4}+2(m_{2}^{T/L})^{2}|\bm{\ell}-\bm{p}|^{2}+(m_{3}^{T/L})^{4}},
\end{equation}
where 
$\bm\ell$ and $\bm p$ respectively denote the quark loop and
the external quark momenta in the DSE,  
$|\bm{\ell}-\bm{p}|=\sqrt{\ell_\parallel^2-2\ell_\parallel p\cos \theta_v +p^2}$, and $\theta_v$ 
is the angle between
$\bm\ell$ and $\bm p$, corresponding to the
polar angle in the angular integration measure~\eqref{eq:misura_angolare_yyy} since $\bm\ell_\parallel$
and $\bm v$ are parallel.

\begin{table*}[t!]
    \centering
    \begin{tabular}{|c|c|c|c|c|c|c|c|c|}
    \hline
       $\mu$ [MeV]  & $Z^T$& $m_1^T$ [MeV] & $m_2^T$ [MeV]& $m_3^T$[MeV] & $Z^L$ & $m_1^L$ [MeV]& $m_2^L$  [MeV] &$m_3^L$ [MeV]\\
       \hline
       400& 1.06518& 1086.16&
          301.016& 621.807& 0.994707& 1376.83 &345.026 & 870.951\\
          500& 0.895304& 1400.91& 268.164& 675.740 &0.922562& 1576.92&338.641&985.201\\
          600& 0.823320 & 1611.01& 220.258 & 709.165& 0.879109 & 1747.78 & 351.646 & 1096.94\\
         \hline
    \end{tabular}
    \caption{Parameters used to fit the gluon propagator with the Gribov-Stingl form~\eqref{gribov-stingl}. }
    \label{tab:params}
\end{table*}

\section{Gap Equation in the CFL phase\label{sec:muoverlabene}}

Next, we turn to the gap equation for the CFL phase, written
in the formalism of HDET.
We define the quark self-energy, $\Sigma$, as
\begin{equation}
  \Sigma= S^{-1}-S_0^{-1}.\
\label{eq:ufficio_1247}
\end{equation}
The the standard one-loop DSE for the quark propagator
can be obained from the effective potential~\eqref{eq:VS_1_lla2}
in the form
\begin{equation}\label{eq:dsesimple2}
\Sigma=\Gamma S \Gamma D\,+ \Gamma^{t}S.
\end{equation}
 
In the color-flavor basis of the CFL phase we can write
Eq.~\eqref{eq:dsesimple2} as 
\begin{eqnarray}
    \Sigma_{AB}(p) &=&-i\frac{4 \pi \mu^2}{(2\pi)^4}\sum_v\int\!\!d^2\ell \Gamma_\mu^{aAC}S_{CD}(\ell)\Gamma_\nu^{bDB}D_{\mu\nu}^{ab}(\ell-p)
\nonumber\\
&&   -i\zeta\frac{4 \pi \mu^2}{(2\pi)^4}\sum_v\int\!\!d^2\ell 
    \Gamma_{ABCD}S_{CD}(\ell)e^{i\theta}  \nonumber\\
    &&+(L\rightarrow R,~\theta\rightarrow-\theta).
    \label{eq:gapequation_1320_2}
\end{eqnarray}
We notice that the change
$\theta\rightarrow-\theta$ is equivalent
to sum up the contributions of the left-handed and
right-handed fields in the OIE term.
$\Sigma_{AB}$ in
Eq.~\eqref{eq:gapequation_1320_2} is still a matrix
in the Nambu-Gorkov space;
taking the $(1,2)$ component of $\Sigma$ in this space
space allows us to extract the gap equation.

\begin{widetext}

Taking into account that $L$ and $R$ are degenerate,
we get
\begin{eqnarray}
\Delta_{AB}(\ell) &=&
i\frac{4\pi\mu^2}{(2\pi)^4}2 g^2\sum_v\int\!\!d^2\ell V_\mu \tilde{V}_\nu D_{\mu\nu}(\ell-p)h_{AaC}h^*_{CaB}\frac{\Delta_C(\ell)}{V\cdot\ell\,\tilde{V}\cdot\ell-\Delta_C^2(\ell)+i 0^+}\nonumber \\
&&+
i\frac{4\pi\mu^2}{(2\pi)^4}2\zeta \cos \theta
\sum_v\int\!\!d\ell^2 \Gamma_{AB}\Gamma_{CD}\frac{\Delta_C(\ell)}{V\cdot\ell\,\tilde{V}\cdot\ell-\Delta_C^2(\ell)+i0^+}\delta_{CD}.
\label{eq:gapequation_1320}
\end{eqnarray}
The overall common coefficient in the two  addenda 
in the right hand side of~\eqref{eq:gapequation_1320}
denotes the HDET integration measure,
see Eq.~\eqref{eq:ufficio_1421},
times the $2$  that accounts for the $L\rightarrow R$ degeneracy. 
The $i0^+$ in the denominators of the
integrand in the above equation correspond to the
standard Feynman prescription to push the negative
energy poles to the upper complex $\ell_0$ plane.

\end{widetext}

For the sake of concreteness, we choose the $A=1,~B=1$
element of $\Sigma$ in Eq.~\eqref{eq:gapequation_1320}.
Performing the summation over the internal color-flavor
indices,
and performing the $\ell_0$ integration via the
residues  we obtain
\begin{eqnarray}
\Delta(p) &=& -\frac{\mu^2 }{6\pi^2}g^2\sum_v\int\!\!d\ell_\parallel V_\mu \tilde{V}_\nu D_{\mu\nu}(\ell-p) \mathcal{W}(\Delta)
\nonumber\\
&&+\frac{2\mu^2}{\pi^2}\zeta \cos \theta
\sum_v
\int\!\!d\ell_\parallel \Delta(\ell_\parallel)\left( 
2\mathcal{X}(\Delta) + \mathcal{X}(2\Delta)
\right).\nonumber\\
&&
\label{eq:gapequafinal_1440}
\end{eqnarray}
Here we defined
\begin{eqnarray}
\mathcal{X}(\Delta) &=&
\frac{1}{\sqrt{\ell_\parallel^2+\Delta(\ell_\parallel)^2}},
\label{eq:xisdefinedas}\\
\mathcal{W}(\Delta) &=&\Delta(\ell_\parallel)\left(
\mathcal{X}(\Delta) + \mathcal{X}(2\Delta)
\right).
\end{eqnarray}
The $V_\mu\tilde V_\nu D_{\mu\nu}(\ell-p)$ in the
equation above can be easily treated by virtue of the
straightforward relations
\begin{eqnarray}
&&P^T_{\mu\nu}(\bm{\ell}-\bm{p})V_\mu \tilde{V}_\nu=|\bm{p}|^2\frac{1-\cos^2\theta_v}{\ell_\parallel^2-2\ell_\parallel |\bm{p}|\cos\theta_v +|\bm{p}|^2},\\
&&P^L_{\mu\nu}(\bm{\ell}-\bm{p})V_\mu \tilde{V}_\nu=1.
\end{eqnarray}
The loop integral involving the OIE term, corresponding to the second term on the right-hand side of Eq.~\eqref{eq:gapequafinal_1440}, diverges for large momenta.  
To regulate this divergence, we introduce a cutoff at $\ell_\parallel=\Lambda$.  
The cutoff value is fixed to $\Lambda=200$ MeV, consistent with previous studies within the HDET framework  \cite{Nardulli:2001iv,Nardulli:2002ma}.

Equation~\eqref{eq:gapequafinal_1440} represents the gap equation in the form we solve numerically.  
Due to the use of a non-local interaction kernel, the gap equation takes the form of a non-linear integral equation for the function $\Delta(\ell)$.  
In the next section, we present some of the numerical solutions of Eq.~\eqref{eq:gapequafinal_1440}.

\section{Solution of the Gap Equation for the CFL Phase\label{sec:ilmilandipioli}}

\begin{figure}[t!]
\includegraphics[totalheight=5cm]{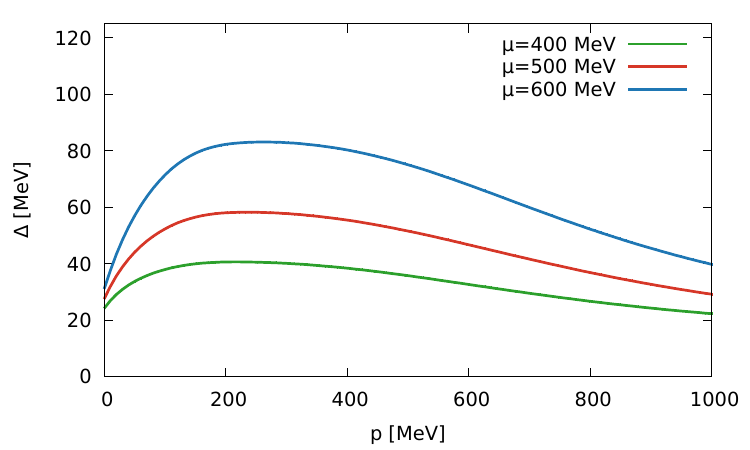}\\
\includegraphics[totalheight=5cm]{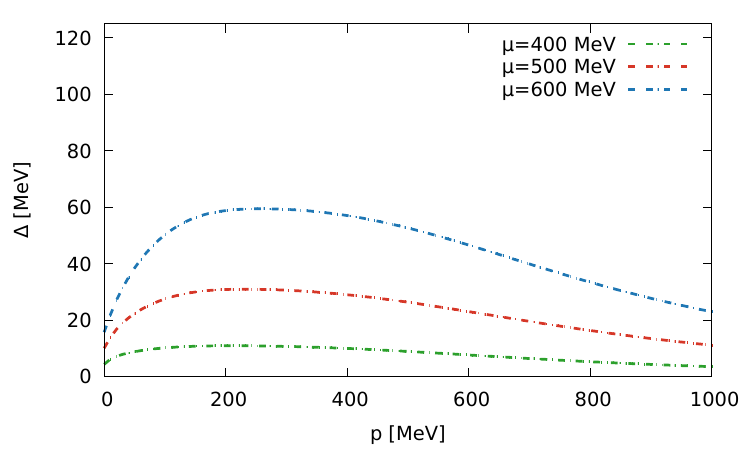}
    \caption{Solution of the gap equation $\Delta$ as a function of the external momentum $p$ in the CFL phase for different values of the  of the chemical potential  $\mu$ at fixed gauge coupling $g=2.8$ and at $\theta=0$. 
    Upper panel corresponds to the solution of the full
    gap equation, while in the lower panel we show the
    results obtained neglecting the OIE term.}
    \label{fig:deltap}
\end{figure}

In this section we present our results for the
solution of the gap equation in the CFL phase.
We used a  fixed-coupling scheme, in which we 
firstly fixed $\alpha_s=g^2/4\pi$ at the scale $\mu$
with the one-loop $\beta$-function, giving
$\alpha_s=1.007,0.762,0.635$ respectively for
$\mu=400,500,600$ MeV. Then, we have
$g=3.558, 3.094, 2.826$ for the same values of $\mu$.
We then fixed $\Lambda=200$ MeV in agreement with previous
calculations in HDET~\cite{Nardulli:2002ma}.
With this choice of parameters we get
$\Delta(p=0)\sim20-40$ MeV 
in the range of $\mu$ considered here,
in agreement with previous studies~\cite{Son:1998uk,Abuki:2002vu,Abuki:2001be, Schafer:1999jg}. 
Finally, we use $M=100$ MeV for 
$\zeta$ in Eq.~\eqref{eq:zeta_coupl_jjj},
which is a fair averaged value of the light and strange quarks
masses,
in agreement with
estimates of the constituent quark masses within 
Nambu-Jona-Lasinio models~\cite{Buballa:2003qv,Ruester:2005jc,Blaschke:2005uj}.
The use of an average value of the constituent quark mass
is justified by the fact that 
$\zeta$ in Eq.~\eqref{eq:zeta_coupl_jjj}
is actually proportional to the trace of the 
quark mass matrix.

In \cref{fig:deltap} we plot the superconductive gap
solution of the gap equation~\eqref{eq:gapequafinal_1440} versus 
the external momentum $p$, for  $\mu=400, 500$ and $600$ MeV. 
In the
upper panel of the figure we show the solution of the full
gap equation, while in the lower panel we plot the
results obtained neglecting the OIE term in the gap equation. 
Firstly, we notice that overall the 
contribution of the latter interaction to the gap is substantial.
Then, we notice that 
for a given value of $\mu$,
our solution shows a maximum
at  $p\neq0$, in  qualitative agreement 
with previous results based on one-gluon-exchange models~\cite{Abuki:2002vu,Abuki:2001be}.
The presence of a maximum for $p\neq 0$ is specific to the gluon propagator used in the gap equation.
Furthermore, as expected from our calculations and in agreement with results known in the literature \cite{Son:1998uk,Abuki:2001be, Schafer:1999jg}, the value of the gap increases as the chemical potential is increased.

We denote by $\Delta_t$ the solution of the full gap equation~\eqref{eq:gapequafinal_1440} in the limit $p\rightarrow\infty$.  
For large external momentum $p\rightarrow\infty$, the gap equation receives contributions only from the contact OIE term, as the term associated with one-gluon exchange on the right-hand side of the equation vanishes due to the asymptotic behavior of the gluon propagator.
From Eq.~\eqref{eq:gapequafinal_1440}
we get
\begin{eqnarray}
\Delta_t &=& \frac{2\mu^2}{\pi^2}\zeta \cos \theta
\sum_v
\int\!\!d\ell_\parallel \Delta(\ell_\parallel)\left( 
2\mathcal{X}(\Delta)+\mathcal{X}(2\Delta)
\right),\nonumber\\
&&
\label{eq:gapequafinal_1440_EB}
\end{eqnarray}
where $\Delta(\ell_\parallel)$ denotes the solution
of the gap equation in the whole momentum range,
and $\mathcal{X}$ is defined in Eq.~\eqref{eq:xisdefinedas}.
The relation~\eqref{eq:gapequafinal_1440_EB} offers an analytical relation between
the integrals on the right-hand side of the equation and
$\Delta_t$, that we will use to obtain a compact expression 
of $\chi$ in the next section.

\begin{figure}[t!]
    \includegraphics[width=\linewidth]{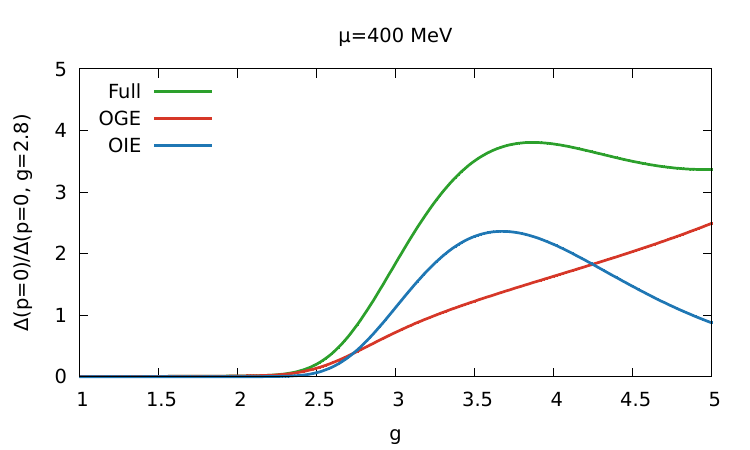}\\
    \includegraphics[width=\linewidth]{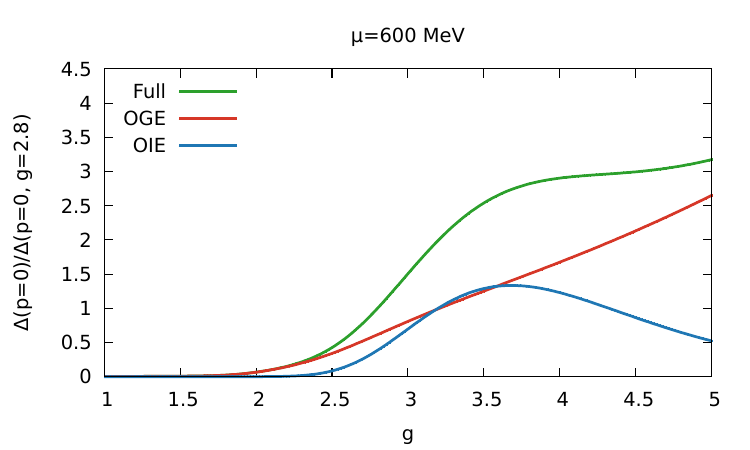}
\caption{Solution of the gap equation $\Delta$  evaluated at vanishing momentum $p=0$ versus the OGE coupling, $g$, at $\theta=0$,  for $\mu=400$ MeV (upper panel) and $\mu=600$ MeV (lower panel), in the CFL phase 
of high-density QCD. For each value of the chemical potential, the curves are normalized to the value of the solution of the gap equation at vanishing momentum $\Delta(p=0)$  for $g=2.8$. }
\label{fig:delta_vs_g}
\end{figure}

It is interesting to analyze the contribution of the
OGE and OIE interactions to the CFL gap. This is shown
in Fig.~\ref{fig:delta_vs_g}, where we plot the solution of the
full gap equation at $p=0$ versus the coupling $g$
for $\mu=400$ MeV (upper panel) and $\mu=600$ MeV
(lower panel). In the two plots, we split the 
contributions of the OGE and OIE to the gap,
which are obtained as the two addenda of the
right hand side of Eq.~\eqref{eq:gapequation_1320}
in which we substitute the solution of the gap equation.
We notice that the relative contribution of the two
interactions to the gap is very sensitive of the 
value of the chemical potential considered, as well as
of $g$. For $g=2.8$ and $\mu=600$ MeV,
the contribution of the two interactions is similar in 
magnitude. On the other hand, for $\mu=600$ MeV and
the same value of $g$, the OIE interaction gives a
relatively smaller contribution. 
These considerations help to extract the dependence of
the topological susceptibility on the couplings 
in Section~\ref{sec:limitcases}.

\begin{figure}[t!]
    \centering
    \includegraphics[width=\linewidth]{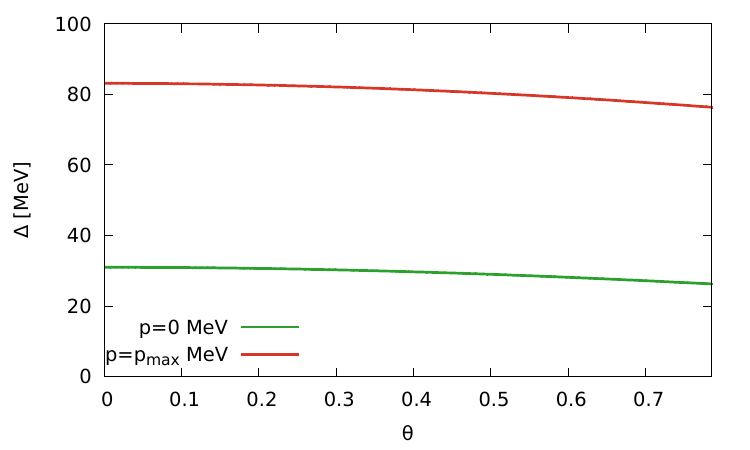}
    \caption{Solution of the gap equation $\Delta$ as a function of $\theta$ at fixed momenta for fixed gauge coupling $g=2.8$ and $\mu=600$ MeV, in the CFL phase. $\Delta$ is evaluated at vanishing momentum and at $p_{max}$ for which $\Delta(p;\theta=0)$ reaches its maximum (see \cref{fig:deltap}).}
\label{fig:eDelta_vs_theta_vari_p}
\end{figure}

\section{Topological susceptibility in the CFL phase}

The main goal of this study is the computation of the
topological susceptibility. 
Identifying the CJT effective potential, $V$, with 
the thermodynamic potential $\Omega$,
we obtain an expression for $\chi$ starting from $V$, that is
\begin{equation}
    \chi=\frac{d^2 V(\Delta(p;\theta), \theta)}{d \theta^2}\Big|_{\theta=0}.
    \label{eq:trtttpp}
\end{equation}
In~\eqref{eq:trtttpp} we denoted by $\Delta(p;\theta)$
the solution of the gap equation, that depends 
on momentum as well as on $\theta$.

In principle, one should take the total derivative
of the thermodynamic potential with respect to
$\theta$, hence considering both the explicit and the implicit
$\theta$-dependence of $V$.
This was the case for the calculation of the topological
susceptibility at $\mu=0$~\cite{Lu:2018ukl,Zhang:2025lan}.
However, our theory remains invariant under the transformation $\theta\rightarrow -\theta$.  
In contrast to the case of chiral and $\eta$-condensates at $\mu=0$, where both scalar and pseudo-scalar condensates exist for $\theta\neq 0$—even for small $\theta$—and $\partial\eta/\partial\theta\neq 0$~\cite{Zhang:2025lan}, in superconducting phases, we find only the scalar condensate~\cite{Murgana:2024djt}, unless $\theta\ge\pi/2$.  
As a result, the effective potential is an even function of $\theta$.  
This implies that every real observable, including the gap $\Delta$, must also be an even function of $\theta$, hence $ \partial \Delta/ \partial\theta=0$ at $\theta=0$.  
This behavior is confirmed by the solution of the 
full gap equation at finite $\theta$, see~\cref{fig:eDelta_vs_theta_vari_p},
in which we notice that $\Delta$ versus $\theta$
is flat for $\theta\rightarrow 0$.  
Using this property, a straightforward application of the chain rule shows that, when computing $\chi$, we only need to consider the explicit dependence of the effective potential on $\theta$.  
The only explicit dependence of $V$ on $\theta$ appears in $V_2^\mathrm{OIE}$ in Eq.~\eqref{eq:ipotenusaecateti}.  
Therefore, instead of Eq.~\eqref{eq:trtttpp}, we use  
\begin{equation}
    \chi=
    \left.\frac{\partial^2 V_2^\mathrm{OIE}(\Delta(p;\theta), \theta)}{\partial \theta^2}\right|_{\theta=0}.
    \label{eq:trtttpp?2}
\end{equation}

In the HDET formulation, we 
can write Eq.~\eqref{eq:ipotenusaecateti} as
\begin{equation}
V_2^\mathrm{OIE}= \frac{1}{2}\frac{4\mu^4}{(2\pi)^6}\zeta\sum_{v,v'} \int d^2\ell d^2k\, \mathrm{Tr}\left[\mathcal{R}(\ell,k) e^{i\theta} \right]~+h.c.,
\label{eq:nefisicanematematicaneingegneria}
\end{equation}
where 
\begin{equation}
\mathcal{R}(\ell,k) =
S_{AB}(\ell)\Gamma_{ABCD}S_{CD}(k).
\label{eq:meno16}
\end{equation}
In Eq.~\eqref{eq:nefisicanematematicaneingegneria}
$\bm v$ and $\bm v^\prime$
denote the Fermi velocities related to the loops in
$\ell$ and $k$ respectively.
The result~\eqref{eq:nefisicanematematicaneingegneria} 
stands for any color-superconductive phase, because we have not
made any assumption on the quark propagator.
We now specialize it to the case of the CFL phase.
Taking into account Eqs.~\eqref{eq:gamma_def_ABCD}
and~\eqref{eq:aldottorato}, with $S_A$ given by 
Eq.~\eqref{eq:s},
we obtain
\begin{equation}
V_2^\mathrm{OIE}=  \frac{\mu^4}{(2\pi)^6}\zeta \cos\theta \int d^2\ell d^2k\, \mathrm{Tr}\left[S_{A}(\ell)\Gamma_{AA} \Gamma_{CC}S_{C}(k)\right],
\label{eq:openuniversity_deivideo_V2t}
\end{equation}
from which we derive
\begin{equation}
\chi=  -\frac{\mu^4}{(2\pi)^6}\zeta  \int d^2\ell d^2k\, \mathrm{Tr}\left[S_{A}(\ell)\Gamma_{AA} \Gamma_{CC}S_{C}(k)\right].
\label{eq:openuniversity_deivideo}
\end{equation}

A straightforward calculation of the trace 
and of the integral over $ \ell_0$ by
residues leads at
\begin{eqnarray}
\chi &=& \frac{\mu^4}{2\pi^4}\zeta
\left[2 \int d\ell_\parallel \frac{\Delta(\ell_\parallel, \theta=0)}{\sqrt{\ell_\parallel^2+\Delta(\ell_\parallel^2, \theta=0)}} \right.\nonumber\\
&&\left.
+ \int d\ell_\parallel \frac{\Delta(\ell_\parallel, \theta=0)}{\sqrt{\ell_\parallel^2+4\Delta(\ell_\parallel^2, \theta=0)}} \right]^2.\label{eq:sustopcfl}
\end{eqnarray}
We notice that the two addenda in the square bracket
in Eq.~\eqref{eq:sustopcfl}
can be directly related to the solution of the gap
equation for $p\rightarrow\infty$ and $\theta=0$, $\Delta_t$, 
see Eq.~\eqref{eq:gapequafinal_1440_EB}.
We therefore get
\begin{equation}
 \chi=\frac{\Delta_t^2}{2\zeta}.
 \label{eq:topol_sus_cfl_tot_1155}
\end{equation}
This is one of the main results of this article:
it expresses the topological susceptibility of the CFL phase
in terms of the gap in the large-momentum limit,
and the value of the  OIE coupling $\zeta$.

We also notice that 
the functional dependence of $\chi$ on
$\zeta$ and $\Delta_t$ in
Eq.~\eqref{eq:topol_sus_cfl_tot_1155}
stands regardless of the gluon propagator used in the 
DSE for the quark self-energy: the former can only
modify the momentum-dependence of the superconductive
gap, 
and consequently the numerical value of $\Delta_t$,
but does not directly affect Eqs.~\eqref{eq:sustopcfl}
and~\eqref{eq:gapequafinal_1440_EB}.

\begin{figure}[t!]
    \includegraphics[totalheight=5cm]{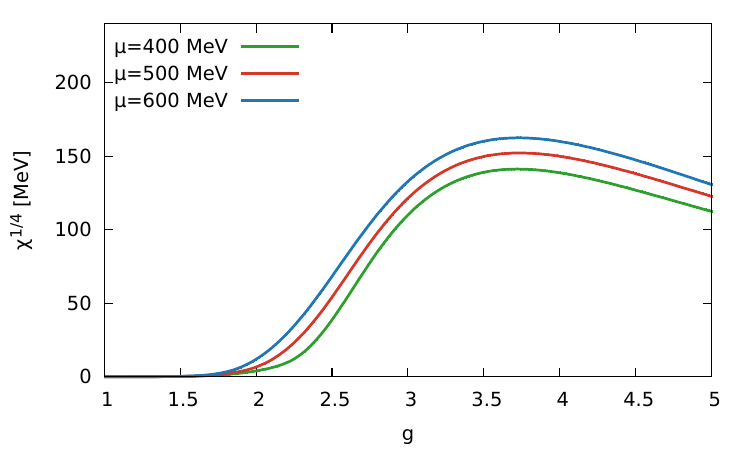}
\caption{$\chi^{1/4}$ as a function of the OGE
coupling, $g$, for several values of $\mu$, 
in the CFL phase. 
}
    \label{fig:sustopmu}
\end{figure}

In order to give concrete numerical values of $\chi$
obtained within our calculation, we quote the result
at $\mu=600$ and $g=2.8$, 
that corresponds to the solution of the
gap equation $\Delta_t=11.9$ MeV:
we find
\begin{equation}
    \chi^{1/4}=111\mbox{ MeV},~~~\mu=600~\mathrm{MeV}.
 \label{eq:volevoiltutor_1158}   
\end{equation}
Similarly, for $\mu=500$ MeV we find $\Delta_t=12.3$ MeV and
\begin{equation}
    \chi^{1/4}=98\mbox{ MeV},~~~\mu=500~\mathrm{MeV}.
 \label{eq:volevoiltutor_1158_500}   
\end{equation}
This number is of the same order of magnitude of
$\chi$ obtained in~\cite{Murgana:2024djt} for the 2SC phase,
which sits in the range $50-75$ MeV for $\mu=400$ MeV
and $\Delta= 25$ MeV. 

In Fig.~\ref{fig:sustopmu} we plot
$\chi^{1/4}$ versus the OGE
coupling, $g$, for several values of $\mu$ 
in the CFL phase. 
Qualitatively, $\chi$ increases with $g$,
due to the increase of $\Delta_t$. 
We notice a strong dependence
of $\chi$ on $g$, which we extract  below
in two limiting cases.
One of the messages encoded in the figure is that
that an uncertainty on $g$ in the CFL phase results in a 
substantial uncertainty on $ \zeta$  and $\chi$.

\section{Topological Susceptibility in the 2SC Phase}

\begin{figure}[t!]
    \includegraphics[totalheight=5cm]{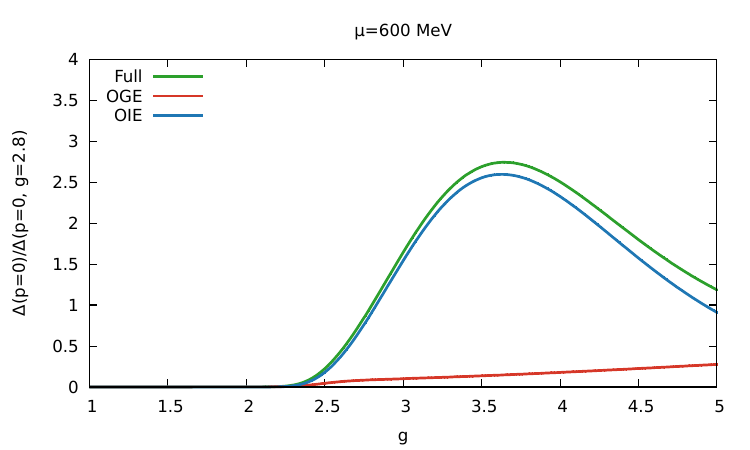}
\caption{Solution of the gap equation $\Delta$  evaluated at vanishing momentum $p=0$ versus the OGE coupling, $g$, at $\theta=0$,  for $\mu=600$ MeV in the 2SC phase 
of high-density QCD. The curves are normalized to the value of the solution of the gap equation at vanishing momentum $\Delta(p=0)$  for $g=2.8$. }
\label{fig:delta_2sc_vs_g}
\end{figure}

Here we briefly extend our previous discussion to the
2SC phase.
In this phase, the quark-quark condensate is
given by \cite{Schafer:1999jg,Abuki:2001be}
\begin{equation}
\langle\psi^{LT}_{\alpha i}C\psi^{L}_{\beta j}\rangle=-\langle\psi^{RT}_{\alpha i}C\psi^{R}_{\beta j}\rangle
\propto\frac{\Delta}{2}\epsilon_{\alpha\beta 3}\epsilon_{i j 3},
\label{eq:calciodirigorenondato_2SC}
\end{equation}
instead of Eq.~\eqref{eq:calciodirigorenondato_2SC}.
Hence, the condensate in the 2SC phase involves 
$u$ and $d$ quarks only, with colors red and green. 
The $U(1)_A$-breaking term has the same form as in
Eq.~\eqref{eq:tandemst2}, with 
\begin{equation}
\zeta=  \int d\rho \, n_0(\rho) \,\left( \frac{4}{3\pi^2 \rho^3}\right)^2.
\label{eq:viakal}
\end{equation}

Analogously to the CFL case, the first step is to solve the gap equation. It is possible to formulate the quark propagator and the whole DSE using a  change in the color-flavor basis which is completely analogous to \cref{eq:moltariserva_1101} (for more details one can refer to \cite{Nardulli:2001iv}). Following straightforward steps, analogous to the ones that leads to Eq.~\eqref{eq:gapequafinal_1440}, one can easily obtain the gap equation for the 2SC phase
\begin{eqnarray}\label{eq:gap2sc}
\Delta(p)&=&-\frac{\mu^2}{3(2\pi)^2}g^2\sum_{\bm{v}}\int\!\!d\ell_\parallel  V_\mu\tilde{V}_\nu D_{\mu\nu}(\ell-p)
 \frac{\Delta(\ell_\parallel)}{\sqrt{\ell_\parallel^2+\Delta^2(\ell_\parallel)}} \nonumber\\
&&+ \frac{ 2\mu^2}{\pi^2}\zeta \cos \theta\sum_{\bm{v}}\int\!\!d\ell_\parallel \frac{\Delta(\ell_\parallel)}{\sqrt{\ell_\parallel^2+\Delta(\ell_\parallel)^2}}.
\end{eqnarray}
As for the CFL case, we can then define the contribution to the gap equation which arises from the OIE vertex considering the $p\to \infty$ limit 
\begin{equation}\label{eq:T'Hooft_2sc}
    \Delta_t=\frac{ \mu^2}{\pi^2}\zeta \cos \theta\int\!\!d\ell_\parallel \frac{\Delta(\ell_\parallel)}{\sqrt{\ell_\parallel^2+\Delta(\ell_\parallel)^2}}.
\end{equation}

Armed with the solution of the gap equation, one can proceed analogously to the steps that led to~\eqref{eq:sustopcfl} and obtain the expression for the topological sucseptibility in the 2SC phase, that is
\begin{equation}\label{eq:sustop2sc}
\chi= \frac{\mu^4}{2\pi^4}\zeta\left(\int d\ell_\parallel \frac{\Delta(\ell_\parallel, \theta=0)}{\sqrt{\ell_\parallel^2+\Delta^2(\ell_\parallel, \theta=0)}}\right)^2.
\end{equation}
We notice that, combining \cref{eq:T'Hooft_2sc} and \cref{eq:sustop2sc} we obtain
\begin{equation}
 \chi=\frac{\Delta_t^2}{2\zeta}.
 \label{eq:topol_sus_cfl_tot_1155_alp}
\end{equation}
that formally coincides with
\cref{eq:topol_sus_cfl_tot_1155}. 
We hence find that the topological susceptibility in the
2SC and the CFL phases depends in the same fashion on 
$\zeta$ and $\Delta_t$: the numerical differences between
the values of $\chi$ in the two phases is encoded $\Delta_t$
and $\zeta$.

As an indication, for $g=2.8$ and $\mu=600$ MeV,
which are the same values we used for the estimate~\eqref{eq:volevoiltutor_1158} 
of $\chi$ in the CFL phase, in the 2SC phase
we find $\Delta(p=0)=208.8$ MeV, $\Delta_t=190.5$  MeV and
\begin{equation}
    \chi^{1/4}=182~\mathrm{MeV},~~~\mu=600~\mathrm{MeV}.
\end{equation}
Similarly, for $\mu=500$ MeV we find $\Delta(p=0)=205.1$  MeV, $\Delta_t=188.5$  MeV and
\begin{equation}
    \chi^{1/4}=165\mbox{ MeV},~~~\mu=500~\mathrm{MeV}.
 \label{eq:volevoiltutor_1158_500_2SC}   
\end{equation}

The result~\eqref{eq:topol_sus_cfl_tot_1155_alp}
is analogous to the one obtained in~\cite{Murgana:2024djt},
in which $\chi$ was computed within an effective model with
a contact interaction both in the one-gluon-exchange and in the
OIE channels for the 2SC phase, namely
\begin{equation}
\chi=\frac{\Delta^2}{2G_D}\xi\frac{2-\xi}{2+\xi}.
\label{eq:xi_1705}
\end{equation}
Here, $\Delta$ stands for the
solution of the gap equation, that takes contribution
both from the one-gluon-exchange and the OIE
channels (both of them are momentum-independent
in~\cite{Murgana:2024djt}, therefore the one-gluon-exchange
contribution does not vanish as it happens
within our model for $p\rightarrow\infty$).
Moreover,
$G_D$ denotes the strength of the coupling in the one-gluon-exchange channel, and 
$\xi$ corresponds to the ratio between the couplings in the
OIE and OGE channels, that was treated
as a free parameter.
One of the improvements that we bring in the present work
for the calculation of $\chi$, in comparison
to~\cite{Murgana:2024djt}, 
are that we adopt a calculation scheme 
based on a DSE for the quark propagator that is more
easily related to full QCD than effective models:
indeed, our results depend solely on the QCD coupling,
$g$, on the strength of the OIE term, which is
computed from the theory of QCD instantons at large $\mu$ 
and is fully specified as soon as
$g$ and $\mu$ are known. Moreover,
we allow for a momentum-dependence of the 
superconductive gap, which is the natural consequence
of the use of an improved gluon propagator in  the kernel
of the gap equation.

\section{Topological susceptibility in limiting cases\label{sec:limitcases}}

As a theoretical exercise, 
we analyze the dependence of $\chi$ on $\zeta$ 
in two limiting cases, assuming, for 
illustrative purposes, that the gap equation is 
dominated either by the OGE or the OIE. 
While such a dominance is not observed in our results, 
this analysis provides insight into the potential behavior 
of the system under extreme assumptions.

To begin with, we analyze $\chi$ in the CFL phase, considering
the case in which the
OGE dominates the CFL gap equation. 
In this regime, on the right-hand side of Eq.~\eqref{eq:gapequafinal_1440_EB}, we can use the solution of Eq.~\eqref{eq:gapequafinal_1440} for $\zeta=0$, leading to $\Delta_t\propto \zeta$ in the weak coupling limit.  
This implies
\begin{equation}
\chi\approx \zeta\frac{\mu^4}{2\pi^4} \mathcal{Y},
\label{eq:inter_troppo_brava}
\end{equation}
where we put
\begin{eqnarray}\label{eq:diventaanchegradevole}
\mathcal{Y} &=& \left[2 \int d\ell_\parallel \frac{\Delta(\ell_\parallel, \theta=0)_{\mathrm{oge}}}{\sqrt{\ell_\parallel^2+\Delta(\ell_\parallel^2, \theta=0)_{\mathrm{oge}}}}\right. + \nonumber\\
&&
\left.\int d\ell_\parallel \frac{\Delta^2(\ell_\parallel, \theta=0)_{\mathrm{oge}}}{\sqrt{\ell_\parallel^2+4\Delta^2(\ell_\parallel^2, \theta=0)_{\mathrm{oge}}}} \right]^2,
\end{eqnarray}
and $\Delta(\ell_\parallel^2, \theta=0)_{\mathrm{oge}}$
corresponds to the solution of the gap
equation for $\zeta=0$.

Next, we turn to the case in which the  OIE contribution
dominates the gap equation.
In the CFL phase this 
In this case,
we can approximate $\Delta(\ell_\parallel)\approx \Delta_t$
for any $\ell_\parallel$. 
In this limit, Eq.~\eqref{eq:gapequafinal_1440_EB}
for $\theta=0$
turns to
\begin{eqnarray}
1 &=& \frac{2\mu^2}{\pi^2}\zeta 
\sum_v
\int\!\!d\ell_\parallel  \left( \frac{2}{\sqrt{\ell_\parallel^2+\Delta_t^2}}+\frac{1}{\sqrt{\ell_\parallel^2+4\Delta_t^2}}\right),
\label{eq:gapequafinal_1440_EB_acerbiinuscita}
\end{eqnarray}
which looks like the standard BCS gap equation at $T=0$
for a contact four-fermion interaction.
The integral on the right hand side of 
Eq.~\eqref{eq:gapequafinal_1440_EB_acerbiinuscita} 
can be computed as in the BCS theory, namely restricting the
integration domain to a shell around the Fermi surface
(corresponding to $\ell_\parallel=0$ in HDET) with
width $\delta$. Assuming $\delta\gg\Delta_t$ we get
\begin{equation}
\Delta_t = \delta \exp\left(-\frac{\pi^2}{6\zeta\mu^2}\right),
\label{eq:ungrandetalento}
\end{equation}
that is, $\Delta_t$ increases with $\zeta$.
As a consequence, 
\begin{equation}
\chi \approx \frac{\delta^2}{2\zeta}
\exp\left(-\frac{\pi^2}{3\zeta\mu^2}\right).
\label{eq:unbuondifensore}
\end{equation}

For the 2SC phase we can repeat the same arguments that
lead at Eqs.~\eqref{eq:inter_troppo_brava}
and~\eqref{eq:unbuondifensore}. In the 
limiting case in which the OGE is more important, 
$\chi$ is given by Eq.~\eqref{eq:inter_troppo_brava} 
with
\begin{eqnarray}\label{eq:biopresto}
\mathcal{Y} &=& \left[\int d\ell_\parallel \frac{\Delta(\ell_\parallel, \theta=0)_{\mathrm{oge}}}{\sqrt{\ell_\parallel^2+\Delta(\ell_\parallel^2, \theta=0)_{\mathrm{oge}}}}\right]^2.
\end{eqnarray}
In the limit in which the OIE dominates the gap,
instead of Eq.~\eqref{eq:gapequafinal_1440_EB_acerbiinuscita}
we get from Eq.~\eqref{eq:T'Hooft_2sc}
\begin{eqnarray}
1 &=& \frac{\mu^2}{\pi^2}\zeta 
\sum_v
\int\!\!d\ell_\parallel  \frac{1}{\sqrt{\ell_\parallel^2+\Delta_t^2}},
\label{eq:baristatec}
\end{eqnarray}
which gives
\begin{equation}
\Delta_t = 2\delta \exp\left(-\frac{\pi^2}{\zeta\mu^2}\right),
\label{eq:ungrandetalento2sc}
\end{equation}
instead of Eq.~\eqref{eq:ungrandetalento}.
Consequently, 
\begin{equation}
\chi \approx \frac{2\delta^2}{\zeta}
\exp\left(-\frac{2\pi^2}{\zeta\mu^2}\right).
\label{eq:unbuondifensore3sc}
\end{equation}
instead of Eq.~\eqref{eq:unbuondifensore}.

\section{Comments on the axion mass}

The calculation of the effective potential in Eq.~\eqref{eq:VS_1_lla2} allows us to directly access the low-energy properties of the QCD axion.
In particular, the axion mass, $m_a$, is related to the topological susceptibility $\chi$ via the relation $\chi = m_a^2 f_a^2$, where $f_a$ is the axion decay constant.
From Eqs.~\eqref{eq:topol_sus_cfl_tot_1155}
and~\eqref{eq:topol_sus_cfl_tot_1155_alp} we obtain
\begin{equation}
m_a^2 = \frac{\Delta_t^2}{2\zeta f_a^2}
 \label{eq:topol_sus_cfl_tot_1155_axionmass}
\end{equation}
for the CFL and the 2SC phases.
Since $f_a$ is not a QCD scale, it is unlikely to be modified during the transition from nuclear matter to quark matter.
Moreover, it cannot be computed within our framework. However, constraints on this quantity are available from various astrophysical and cosmological observations
(see~\cite{PDG2024,DiLuzio2021,Carenza2024} for astrophysical bounds).

Within our formalism, Eq.~\eqref{eq:topol_sus_cfl_tot_1155_axionmass} is exact and allows us to express the axion mass in the CFL and 2SC phases of QCD solely in terms of the superconducting gap and the OIE coupling.
The details of the QCD interaction leading to color superconductivity are encapsulated in the value of $\Delta_t$ and do not alter the analytical dependence of $m_a$ on the gap. 

Our result, Eq.~\eqref{eq:topol_sus_cfl_tot_1155_axionmass}, can be compared with previous works, such as~\cite{Murgana:2024djt}, where it was found that
\begin{equation}
m_a^2 = \frac{\Delta^2}{2G_D f_a^2} \xi \frac{2-\xi}{2+\xi}.
\label{eq:xi_1705_axionmass}
\end{equation}
The improvement provided by Eq.~\eqref{eq:topol_sus_cfl_tot_1155_axionmass} over Eq.~\eqref{eq:xi_1705_axionmass} is that $m_a$ in Eq.~\eqref{eq:topol_sus_cfl_tot_1155_axionmass} is expressed purely in terms of QCD parameters, rather than in terms of parameters from effective models.
Apart from the relation with the OIE coupling, which is computed at large $\mu$ and expressed solely in terms of $g$, all details of the gap equation kernel are embedded in $\Delta_t$.

We find that $\chi$ in the superconducting phases is of the same order as in the vacuum. 
Therefore, one of the consequences of our work is that the axion mass in these phases can be as large as at $\mu = 0$, provided that phase transitions to color-superconducting phases occur in dense matter.
This result 
is in agreement with~\cite{Murgana:2024djt,Zhang:2025lan}
but differs from that in~\cite{Zhang:2023lij}, where color superconductivity was not included, and the only phase transition considered was that to normal quark matter.
Thus, if the transition to the color-superconducting phase at large $\mu$ is preceded, at lower $\mu$, by a normal quark matter phase, then we expect a decrease in $m_a$ in the latter phase, followed by an increase in the former.
On the other hand, if the transition from nuclear matter to quark matter occurs without an intermediate-$\mu$ phase,
then $m_a$ might not undergo a significant change from its vacuum value.

It is certainly interesting to extend these considerations to other low-energy properties of the QCD axion, {\it in primis}, the axion self-coupling.
This investigation will be pursued in future works.

\section{Conclusions and outlook}

We computed the topological susceptibility, $\chi$,
in the color-superconductive phases of Quantum Chromodynamics
(QCD) at zero temperature and large quark chemical
potential, $\mu$.
We focused on the Color-Flavor-Locking (CFL) and the
2-Color-Superconductor (2SC) phases.
For the CFL, we assume the quarks have an average 
constituent mass $M$, whose value is borrowed from the
results of NJL-like models at finite $\mu$,
while for the 2SC phase we assume massles $u$ and $d$ quarks.

Our calculation scheme combines for the first time
the High Density Effective Theory (HDET) 
at finite $\theta$
to the
2-particle irreducible formalism, that we use to derive
the Dyson-Schwinger equation (DSE) for the quark propagator
in the rainbow approximation (bare vertices)
and fixed-coupling.
In the kernel of the DSE we use both an improved
gluon propagator and a local  vertex, 
that respectively describe
the contributions of the 
one-gluon-exchange (OGE)
and of the one-instanton-exchange (OIE) 
interactions to the supercondutcive gap.
For the sake of simplicity, we work in the
static limit of the gluon propagator, which implies
that the gap function, $\Delta$, does not depend on energy.
The gap equation in this formalism leads to a 
non-linear integral equation, that we solve numerically.

Within our calculation scheme,
for the CFL phase
there are only three free
parameters in the quark sector, that are the OGE coupling, $g$, 
the average quark mass, $M$, 
and the cutoff, $\Lambda$, that regulates the divergence
of the OIE loop in the gap equation.
In the 2SC phase, the quark masses are assumed to vanish,
so we are left with $g$ and $\Lambda$ as free parameters.
We fixed $g$ at the scale $\mu$ via the one-loop
QCD $\beta$-function.
The value of $\Lambda$ has been fixed in agreement with 
previous calculations
in HDET~\cite{Nardulli:2002ma}. 
The strength of the OIE coupling, $\zeta$,
has been fixed by the perturbative results at finite $\mu$.
The gluon propagator was instead
borrowed from~\cite{Comitini:2024xjh},
and entails non-perturbative mass scales
arising both from vacuum and from matter effects.
We regularized the theory via the introduction of the cutoff $\Lambda$.

We derived an analytical result for the topological
susceptibility in the CFL and 2SC phases, see Eq.~\eqref{eq:topol_sus_cfl_tot_1155}, which stands
regardless of the specific form of the gluon propagator
used in the OGE contribution to the
gap equation. In fact, we were able to express $\chi$
as a function of $\zeta$ and
$\Delta_t$, namely the solution of the gap equation
in the large-$p$ limit.
Our calculations give
values of $\chi$ that are of the same
order of those in the vacuum both for the CFL and for
the 2SC phases. 
This result is in some agreement with previous calculations
based on effective models with a contact interaction~\cite{Murgana:2024djt,Zhang:2025lan},
where however
there is some additional uncertainty on the final prediction
for $\chi$ due to the treatment of the ratio of the couplings
in the OGE and OIE as a free parameter.
We also derived two limiting behaviors of
$\chi$ versus $\zeta$, for the cases in which
the gap equation
is dominated either by the OGE or by the OIE kernels,
highlighting a linear increase of $\chi$ in the former case,
and a BCS-like dependence of $\chi$ in $\zeta$ in the
latter case, see Eqs.~\eqref{eq:inter_troppo_brava}
and~\eqref{eq:unbuondifensore} respectively.

Finally, we related our findings for $\chi$ to 
the QCD-axion mass, $m_a$.
In fact,
we propose Eq.~\eqref{eq:topol_sus_cfl_tot_1155_axionmass}
as a relation between $m_a$, the superconductive gap
and $\zeta$, which stands both in the CFL and the 2SC phases.
From the fact that the values of $\chi$ 
in the superconductive phases
are in the same ballpark of those in the vacuum,
suggests that $m_a$ in superdense QCD phases
migh be as large as that in the vacuum.
This is different from the behavior of $m_a$
found in~\cite{Zhang:2023lij} where color superconductivity
was not considered, and $m_a$ was found to be
quite smaller than its vacuum value.
Hence, if a phase transition happens directly from 
nuclear matter to superconductive quark matter,
as strong-coupling scenarios seem to suggest~\cite{Ruester:2005jc,Blaschke:2005uj}
then $m_a$ at large $\mu$ might not be
very different from its value in the vacuum,
see also~\cite{Zhang:2025lan}.
Moreover, since the topological susceptibility can be
directly related to the surface tension of the axion walls~\cite{Zhang:2023lij,Zhang:2025lan}, our results
suggest that the surface tension of these walls in
superdense QCD phases might be as large as that in the vacuum.

Admittedly, our approach to the DSE is oversimplified
in comparison with a full QCD approach. Indeed, we 
employ an HDET version of the effective potential and of the
gap equation of dense QCD, 
consider the static approximation for the quark and the
gluon propagator, 
ignore the backreaction of color superconductivity
to the gluon propagator (although the latter contains
screening effects, computed non-perturbatively in \cite{Comitini:2024xjh}), and adopt the rainbow
approximation for the quark-gluon vertex.
Nevertheless, ours is a first step to the application 
of Dyson-Schwinger equations in dense QCD to the
calculation of the topological susceptibility,
and paves the way for more refined calculations that remove
the simplifications we have done. 
More work on this subject
is ongoing and we plan to report on it in future works.

\subsection*{Acknowledgments}
M. R. acknowledges Bruno Barbieri, Lautaro Martinez 
and John Petrucci for inspiration.
M. R. acknowledges discussions with Ana G. Grunfeld and
David E. Alvarez Castillo.
This work has been partly funded by 
PIACERI “Linea di intervento 1” (M@uRHIC) of the University of Catania, and 
the European Union – Next Generation EU through the research grants number P2022Z4P4B “SOPHYA - Sustainable Optimised PHYsics Algorithms: fundamental physics to build an advanced society” and 2022SM5YAS, under the program PRIN 2022 PNRR of the Italian Ministero dell’Università e Ricerca (MUR).

\bibliography{topsusc}

\end{document}